\documentclass[fleqn,twoside]{article}
\usepackage{espcrc2}
\usepackage{graphicx}
\newcommand{\I}{\ensuremath{\mathrm{i}\,}}
\newcommand{\E}{\ensuremath{\mathrm{e}\,}}
\title{Chiral perturbation theory for twisted mass QCD}

\author{\underline{Gernot M\"unster}\address[MS]{Institut f\"ur Theoretische
             Physik, Universit\"at M\"unster,
             Wilhelm-Klemm-Str.~9, D-48149 M\"unster, Germany},
        Christian Schmidt\addressmark[MS]
        and
        Enno E.~Scholz\address{Deutsches Elektronen-Synchrotron DESY,
        Notkestr.~85, D-22603 Hamburg, Germany}}

\begin{document}

\begin{abstract}
Quantum Chromodynamics on a lattice with Wilson fermions and a chirally
twisted mass term for two degenerate quark flavours is considered in the
framework of chiral perturbation theory. The pion masses and decay constants
are calculated in next-to-leading order including terms linear in the
lattice spacing $a$. We treat both unquenched and partially quenched QCD. We
also discuss the phase structure of twisted mass lattice QCD.
\vspace{1pc}
\end{abstract}

\maketitle
%
%%%%%%%%%%%%%%%%%%%%%%%%%%%%%%%%%%%%%%%%%%%%%%%%%%%%%%%%%%%%%%%%%%%%%%%%
%
\section{Introduction}
Lattice QCD is fighting on several fronts in order to obtain results
relevant to the physics of hadrons in the continuum: one wants to control
the effects (a) of the finite size of the lattice, (b) of the non-zero
lattice spacing $a$, and (c) of too large quark masses. For small quark
masses Monte Carlo calculations suffer from a slowing down roughly
proportional to $m_q^p$ with $p=2 - 3$. This has so far prevented
simulations with light Wilson quarks of realistic masses.

For the extrapolation of the numerical results to small values of up- and
down-quark masses, chiral perturbation theory can be employed. It amounts to
an expansion around the chiral limit at vanishing quark masses. The low
energy parameters of chiral perturbation theory, the
\textit{Gasser-Leutwyler coefficients}, can in turn be determined by
numerical simulations of lattice QCD, see \cite{Wittig} for a review.
%%%%%%%%%%%%%%%%%%%%%%%%%%%%%%%%%%%%
%
\subsection{Twisted mass lattice QCD}
For simplicity we consider QCD with $N_f = 2$ flavours of degenerate light
quarks, $m_u = m_d$, in the following. Consider a ``twisted'' quark mass
matrix ($\tau_3$ acts in flavour space)
\begin{displaymath}
M(\omega) = m_q \, \E^{\I \omega \gamma_5 \tau_3}
= m + \I \mu \gamma_5 \tau_3\,,
\end{displaymath}
where $m = m_q \cos( \omega), \ \mu= m_q \sin( \omega)$. In the continuum
the angle $\omega$ can be removed by a chiral rotation
$q = \E^{-\I \omega \gamma_5 \tau_3 /2} q'$, so that physics does not depend
on $\omega$. On the lattice, however, chiral symmetry is broken even for
massless QCD, and there is a dependence on $\omega$ due to lattice
artifacts.  Twisted mass lattice QCD has been introduced in \cite{TM}.
Recently it has been advocated to employ a chirally twisted quark mass
matrix for Wilson fermions in order to improve the efficiency of QCD
simulations due to full $\mathcal{O}(a)$ improvement occurring for $\omega =
\pi/2$ \cite{FR1}.
%%%%%%%%%%%%%%%%%%%%%%%%%%%%%%%%%%%%
%
\subsection{Chiral perturbation theory}
The $\textrm{SU}(N_f)_L \otimes \textrm{SU}(N_f)_R$ chiral symmetry of
massless QCD is spontaneously broken to $\textrm{SU}(N_f)_V$, and in
addition explicitly broken by non-vanishing quark masses. For $N_f = 2$ the
corresponding Pseudo-Goldstone bosons are the pions $\pi_b$. Chiral
perturbation theory describes their dynamics by means of a low-energy
Lagrangian. It is expressed in terms of the matrix-valued field
$U(x)=\exp ( \I\,\pi_b(x) \tau_b / F_0 )$, which transforms as
$U \rightarrow L U R^{-1}$ under chiral transformations
$(L,R) \in \textrm{SU}(N_f)_L \otimes \textrm{SU}(N_f)_R$.

The leading order effective Lagrangian is
\begin{displaymath}
\mathcal{L}_2
= \frac{F_0^2}{4}\,\mbox{Tr} \left( \partial_\mu U^\dagger \, \partial^\mu U
\right)
- \frac{F_0^2}{4}\,\mbox{Tr} \left( \chi U^\dagger + U \chi^\dagger
\right),
\end{displaymath}
where the quark masses are contained in the matrix
$\chi = 2 B_0 m \, {\bf 1}$.
In higher orders, terms with more fields and/or derivatives appear, which are
multiplied by the Gasser-Leutwyler coefficients.
%%%%%%%%%%%%%%%%%%%%%%%%%%%%%%%%%%%%
%
\subsection{Chiral perturbation theory for lattice QCD}
Chiral perturbation theory can also be applied to QCD on a lattice, see
Oliver B\"ar's talk at this conference. The fields of the low-energy
effective Lagrangian are still defined in the continuum, but the lattice
artifacts are taken into account by additional terms in $\mathcal{L}$
proportional to powers of $a$ \cite{ShaSi,Rupak-Shoresh}. Physical
quantities, like $m_\pi^2$ and $F_{\pi}$, appear in double expansions in
quark masses (modified by logarithms) and in $a$. In leading order the
lattice terms read $\mathcal{L}_{2L}
= - (F_0^2/4)\,\mbox{Tr}\,( \rho\, U^\dagger + U  \rho^\dagger)$, where
$\rho= 2 W_0 a \, {\bf 1}$, with a new parameter $W_0$. Next-to-leading
order calculations have been done in \cite{Rupak-Shoresh} up to
$\mathcal{O}(a)$ and in \cite{BRS} up to $\mathcal{O}(a^2)$.
%%%%%%%%%%%%%%%%%%%%%%%%%%%%%%%%%%%%%%%%%%%%%%%%%%%%%%%%%%%%%%%%%%%%%%%%
%
\section{Chiral perturbation theory for twisted mass lattice QCD}
In view of Monte Carlo calculations in twisted mass lattice QCD, it is
desirable to extend chiral perturbation theory to this case. This has been
done in \cite{MS}. The twisting of the mass term
$\chi \to \chi( \omega) = 2 B_0 m_q \, \E^{-\I \omega \tau_3}$
can be shifted to the lattice term by a chiral rotation
$\rho \to \rho(\omega) = \E^{\I \omega \tau_3} \rho$.

In leading and next-to-leading order (NLO) the Lagrangian correspond to the
one of \cite{Rupak-Shoresh} with a twisted lattice term $\rho(\omega)$.
Spurion analysis shows that the twisting of the mass term does not produce
further terms.

The vacuum corresponds to the minimum of $\mathcal{L}$. In contrast to the
untwisted case, the minimum of the effective action is not located at
vanishing fields, but at a point where $\pi_3 \neq 0$, displaying an
explicit flavour and parity breaking. Chiral perturbation theory amounts to
an expansion around this shifted vacuum.
%%%%%%%%%%%%%%%%%%%%%%%%%%%%%%%%%%%%
%
\subsection{Pion mass and decay constant}
Expansion of $\mathcal{L}$ in terms of pion fields yields the tree level
contribution to the pion propagator. Loop contributions come from the
leading order vertices. New vertices arising from the shift of the vacuum
yield contributions at order $a^2$ only. As a result we obtain\\[2mm]
$m_{\pi}^2= \chi_0 + \rho_0 + 8 (\chi_0^2/F_0^2)
(4 L_6^r + 2 L_8^r - 2 L_4^r - L_5^r)$\\[1mm]
$+8 (\chi_0 \rho_0/F_0^2)
(4 W_6^r + 2 W_8^r - 2 W_4^r - W_5^r - 2 L_4^r - L_5^r)$\\[1mm]
$+ ((\chi_0+\rho_0)^2/(32 \pi^2 F_0^2))\,
\ln ((\chi_0 + \rho_0)/\Lambda^2)\,,$\\[1mm]
where $\chi_0 = 2 B_0 m_q\,,\ \rho_0 = 2 W_0 a\, \cos (\omega)$.
Here $L_k^r$ and $W_k^r$ are renormalized chiral parameters and $\Lambda$ is
the renormalization scale.

At order $a$, the dependence on the twist angle $\omega$ shows up in factors
$\cos (\omega)$. It should be noted, however, that this is different in
higher orders, where the above mentioned shift in the pion fields introduces
new vertices.

In the case of maximal twist, $\omega = \pi/2$, the lattice artifacts vanish
to order $a$. This has been observed for lattice QCD in general in
\cite{FR1}, and is the basis of the improvement proposal made there.

The second physical quantity we calculated is the pion decay constant
$F_{\pi}$, given by
$\langle 0 | J_A^{\mu,a} | \pi_b(p) \rangle = \I F_{\pi} p^\mu \delta_{ab}$,
where $J_A$ is the axial current. A one-loop calculation gives\\[2mm]
$F_{\pi} = F_0 ( 1 + (4/F_0^2) [ \chi_0 \, (2 L^r_4 + L^r_5)
+ \rho_0 \, (2 W_4^r$\\[1mm]
$+ W_5^r)]  - (1/16\pi^2 F_0^2)
[ (\chi_0 + \rho_0) \ln ((\chi_0 + \rho_0)/\Lambda^2) ]).$
%%%%%%%%%%%%%%%%%%%%%%%%%%%%%%%%%%%%%%%%%%%%%%%%%%%%%%%%%%%%%%%%%%%%%%%%
%
\section{Partially quenched lattice QCD}
Partially quenched QCD is an algorithmic approach to the regime of small
quark masses. The Monte Carlo updates are being made with sea-quarks, which
have large enough masses $m_S$ in order to allow a tolerable simulation
speed.  On the other hand, quark propagators and related observables are
evaluated with smaller valence-quark masses $m_V$. Chiral perturbation
theory has been adopted to the case of partially quenched QCD in
\cite{BG,SharpePQ}.

Combining the approaches mentioned above, it appears attractive to simulate
QCD with chirally twisted quark masses in a partially quenched manner. For
the theoretical analysis of the data the extension of the results of
\cite{MS} to the partially quenched case has been done in \cite{MSS}.

In partially quenched chiral perturbation theory the field $U(x)$ is
extended to a graded matrix in $\textrm{SU}(4|2)$. The quark mass matrix is
in our case given by $M = \mbox{diag}(m_V, m_V, m_S , m_S, m_V, m_V)$.
Its twisted counterpart is
$M(\omega_V,\omega_S) = M \,\E^{\I \omega_V \tau_3^V \gamma_5}\
\E^{\I \omega_S \tau_3^S \gamma_5}\ \E^{\I \omega_V \tau_3^G \gamma_5}$.

We have calculated the pion masses $m_{SS}$, $m_{VV}$, $m_{VS}$ and decay
constants $F_{SS}$, $F_{VV}$ and $F_{VS}$ at NLO (one-loop) including
$\mathcal{O}(a)$ \cite{MSS}. At order $a$ the dependence on the twist angle
amounts to factors $\cos \omega_V$ or $\cos \omega_S$. The expressions can
be used in the analysis of numerical results from Monte Carlo calculations
and will aid the extrapolation to small quark masses.

An extension of the calculation to order $a^2$ is in progress. For the
unquenched case results have been published recently by Scorzato
\cite{Scorzato} and Sharpe and Wu \cite{ShaWu1,ShaWu2}.
%%%%%%%%%%%%%%%%%%%%%%%%%%%%%%%%%%%%%%%%%%%%%%%%%%%%%%%%%%%%%%%%%%%%%%%%
%
\section{Phase structure of twisted mass QCD}
A prerequisite to any numerical simulation project is the knowledge of the
phase structure of the model under consideration. Where are lines or points
of phase transitions and how do physical quantities like particle masses
behave near them? The phase structure of twisted mass lattice QCD has been
discussed in \cite{GM,Scorzato,ShaWu1,ShaWu2} on the basis of chiral
perturbation theory.

For lattice QCD without twist, Aoki has proposed the possibility of a phase
with spontaneous flavour and parity breaking \cite{Aoki}. An analysis of
this scenario based on chiral perturbation theory has been made by Sharpe
and Singleton \cite{ShaSi}. A central role plays the potential
$V = -c_1\, u_0 + c_2\, u_0^2$ contained in the effective Lagrangian, where
$u_0 = \frac{1}{2} \mbox{Tr} (U)$. It contains parameters
$c_1 \sim m_q,\ c_2 = \mathcal{O}(a^2)$ for small $m_q = \mathcal{O}(a^2)$.

In twisted mass QCD, where $\mu = m_q \sin( \omega) \neq 0$, the potential
$V$ gets additional contributions: $V = -c_1\, u_0 + c_2\, u_0^2 + c_3\, u_3$,
where $c_3 = 2 F_0^2 B_0 \, \mu\,, \ 
u_3 = \frac{1}{2\I} \mbox{Tr} (\tau_3 U)$.

Depending on the sign of $c_2$, the possible scenarios are

$c_2 > 0$\,: Aoki scenario near the ``critical hopping parameter''
$\kappa_c$ with explicit flavour and parity breaking, and massive pions.

$c_2 < 0$\,: normal scenario with a 1$^{\mbox{st}}$ order phase transition
extending into the $\mu \neq 0$ region, with a second order end point at
$\mu_c = |c_2|/F_0^2 B_0 \sim a^2$.

On the phase transition line the jump in the quark condensate and the
neutral pion mass decrease to zero, when the endpoint is approached:
$(\Delta <\bar{\chi} \chi>)^2 \sim m_{\pi 3}^2
= (1/2 F_0^2 |c_2|) ( 4 c_2^2 - c_3^2 ) \sim a^2$.

Recent Monte Carlo calculations \cite{QCDtm1,QCDtm2} at $\beta = 5.2$
indicate the presence of the normal scenario with a first order line. Owing
to the associated two-phase coexistence and metastability, this represents a
problem for numerical simulations and it would be desirable to have the
phase transition line as short as possible.
\begin{figure}[ht!]
\vspace*{-10mm}
\includegraphics[width=45mm]{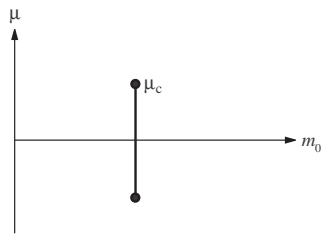}
\vspace*{-10mm}
\caption{Phase structure for the normal scenario}
\vspace*{-11mm}
\end{figure}
%
%%%%%%%%%%%%%%%%%%%%%%%%%%%%%%%%%%%%%%%%%%%%%%%%%%%%%%%%%%%%%%%%%%%%%%%%
%

%
\end{document}